# Temporal and Spatial Evolutions of a Large Sunspot Group and Great Auroral Storms around the Carrington Event in 1859

Hisashi Hayakawa (1-2)*, Yusuke Ebihara (3-4), David M. Willis (2, 5), Shin Toriumi (6), Tomoya Iju (7), Kentaro Hattori (8), Matthew N. Wild (2), Denny M. Oliveira (9-10), Ilaria Ermolli (11), José R. Ribeiro (12), Ana P. Correia (12), Ana I. Ribeiro (13-14), and Delores J. Knipp (15-16)

* hayakawa@kwasan.kyoto-u.ac.jp; hisashi.hayakawa@stfc.ac.uk

(1) Graduate School of Letters, Osaka University, 5600043, Toyonaka, Japan (JSPS Research Fellow).
(2) UK Solar System Data Centre, Space Physics and Operations Division, RAL Space, Science and Technology Facilities Council, Rutherford Appleton Laboratory, Harwell Oxford, Didcot, Oxfordshire, OX11 0QX, UK
(3) Research Institute for Sustainable Humanosphere, Kyoto University, Uji, 6100011, Japan
(4) Unit of Synergetic Studies for Space, Kyoto University, Kyoto, 6068306, Japan
(5) Centre for Fusion, Space and Astrophysics, Department of Physics, University of Warwick, Coventry CV4 7AL, UK
(6) Institute of Space and Astronautical Science (ISAS), Japan Aerospace Exploration Agency (JAXA), 3-1-1 Yoshinodai, Chuo-ku, Sagamihara, Kanagawa 252-5210, Japan
(7) National Astronomical Observatory of Japan, 1818588, Mitaka, Japan.
(8) Graduate School of Science, Kyoto University, Kyoto, Kitashirakawa Oiwake-cho, Sakyo-ku, Kyoto, 6068502, Japan
(9) Goddard Planetary Heliophysics Institute, University of Maryland Baltimore County, 1000 Hilltop Circle Baltimore MD 21250, United States
(10) NASA Goddard Space Flight Center, 8800 Greenbelt Road, Greenbelt MD 21021, United States
(11) INAF Osservatorio Astronomico di Roma, Via Frascati 33 00078 Monte Porzio Catone, Italy
(12) Escola Secundária Henrique Medina, Esposende Av. Dr. Henrique Barros Lima 4740-203 Esposende, Portugal

(13) EPIUnit - Instituto de Saúde Pública, Universidade do Porto Rua das Taipas, 135 4050-600 Porto, Portugal.

(14) Departamento de Ciências da Saúde Pública e Forenses e Educação Médica, Faculdade de Medicina, Universidade do Porto, 4200-319 Porto, Portugal.

(15) High Altitude Observatory, National Center for Atmospheric Research, Boulder, CO, CO 80307, USA

(16) Smead Aerospace Engineering Sciences Department, University of Colorado Boulder, Boulder, CO, CO 80309, USA

**Abstract:**

The Carrington event is considered to be one of the most extreme space weather events in observational history within a series of magnetic storms caused by extreme interplanetary coronal mass ejections (ICMEs) from a large and complex active region (AR) emerged on the solar disk. In this article, we study the temporal and spatial evolutions of the source sunspot active region and visual aurorae, and compare this storm with other extreme space weather events on the basis of their spatial evolution. Sunspot drawings by Schwabe, Secchi, and Carrington describe the position and morphology of the source AR at that time. Visual auroral reports from the Russian Empire, Iberia, Ireland, Oceania, and Japan fill the spatial gap of auroral visibility and revise the time series of auroral visibility in mid to low magnetic latitudes (MLATs). The reconstructed time series is compared with magnetic measurements and shows the correspondence between low to mid latitude aurorae and the phase of magnetic storms. The spatial evolution of the auroral oval is compared with those of other extreme space weather events in 1872, 1909, 1921, and 1989 as well as their storm intensity, and contextualizes the Carrington event, as one of the most extreme space weather events, but likely not unique.

**Plain Language Summary**

The Carrington event is considered to be one of the most extreme space weather events in observational history. In this article, we have studied the temporal and spatial evolutions of the source active region and visual low latitude aurorae. We have also compared this storm with other extreme space weather events on the basis of the spatial evolution. We have compared the available sunspot drawings to reconstruct the morphology and evolution of sunspot groups at that time. We have surveyed visual auroral reports in the Russian Empire, Ireland, Iberian Peninsula, Oceania, and Japan, and fill the spatial gap of auroral visibility and revised its time series. We have compared this time series with magnetic measurements and shown the correspondence between low to mid latitude aurorae and the phase of magnetic storms. We have compared the spatial evolution of the auroral oval with those of other extreme space weather events in 1872, 1909, 1921, and 1989 as well as their storm intensity, and concluded that the Carrington event is one of the most extreme space weather events, but is likely not unique.

**Key Points**
1) Original sunspot drawings during the 1859 storms are revealed and analyzed
2) New auroral reports from Eurasia and Oceania fill the spatial and temporal gaps of the auroral visibility during the 1859 storms
3) The 1859 storms are compared and contextualized with the other extreme space weather events

**1. Introduction:**
After the earliest datable observation of a white-light flare in a large sunspot group by Carrington (1859) and Hodgson (1859) on 1859 September 1, humanity experienced one of the most extreme magnetic storms in observational history (Tsurutani et al., 2003; Cliver and Dietrich, 2013). The reported white-light solar flare was followed by a sudden ionospheric disturbance, namely a large magnetic crochet ≈ 110 nT (Stewart, 1861; Boteler, 2006), which suggests the flare intensity as ≈ X45 – one of the largest in observational history and comparable to the largest modern flare on 2003 November 4 (Boteler, 2006; Cliver and Dietrich, 2013; Curto et al., 2016; *c.f.*, Thomson et al., 2004).

The Carrington event has been thus considered a benchmark of extreme space weather events in terms of its sudden ionospheric disturbance, solar energetic particle (SEP), solar wind velocity, magnetic disturbance, and equatorward boundary of auroral display (Cliver and Svalgaard, 2004). We note that recent discussions on the ice core data (*e.g.*, Wolff et al., 2012; Usoskin and Kovaltsov, 2012; Schrijver et al., 2012; Mekhaldi et al., 2018) made the existing estimate of its SEP fluence (*e.g.*, McCracken et al., 2001; Shea et al., 2006; Smart et al., 2006) rather controversial (Cliver and Dietrich, 2013; Usoskin, 2017).

This great magnetic storm is characterized by an extreme negative magnetic excursion. For example, a value of ≈ −1600 nT was measured at Bombay (N18°56′, E072°50′) (Tsurutani et al., 2003). The anomalously short duration of the negative magnetic excursion has attracted much attention with regard to its cause in relation to the equatorward boundary of auroral visibility, ≈ 23° (Tsurutani et al., 2003) vs ≈ 18° (Green and Boardsen, 2006) in magnetic latitude (MLAT). The current source of the large-amplitude magnetic disturbance is also a matter of controversy. One possible source is the enhanced ring current (Tsurutani et al., 2003; Li et al., 2006; Keika et al., 2015; *c.f.*, Daglis et al., 1999), one is the auroral electrojet (Akasofu and Kamide, 2005; Green and Boardsen, 2006; Cliver and Dietrich, 2013), and another is field aligned current (Cid et al., 2015).

Another characteristic of this event was the great auroral displays down to mid- to low-magnetic latitudes (*e.g.*, Green and Boardsen, 2006; Ribeiro et al., 2011; Hayakawa et al., 2018b; 2018c). The temporal and spatial evolution of auroral visibility was compared with the location of the magnetic record at Bombay by Green and Boardsen (2006). Their auroral records were concentrated in the western hemisphere. The equatorward boundary of the auroral oval reconstructed from the contemporary observational reports was as low as ≈ 28.5°/30.8° invariant latitude (ILAT) (Hayakawa et al., 2018b). Note that ILAT signifies a parameter for the magnetic field line, along which electron moves and cause auroral brightening (O'Brien et al., 1962; Hayakawa et al., 2018b).

The auroral visibility around the Eurasian Continent remains largely unexamined, except for the records in Western Europe and East Asia (Green and Boardsen, 2006; Hayakawa et al., 2018b). Since the magnetic disturbances depend on magnetic local time, the simultaneous observations of the magnetic disturbances and the auroral visibilities provide a better understanding of the Carrington event. Thus, it is of significant interest to revise the temporal and spatial evolution of the auroral oval during the stormy interval around the Carrington event, on the basis of the uncovered contemporary observational reports around the Eurasian Continent, and compare them with the magnetic observations at the time.

The survey on spatial evolution of auroral oval benefits comparison of intensity of extreme space weather events, due to the empirical correlation between equatorward boundary of auroral oval and storm intensity in Dst index (Yokoyama et al., 1998). While the Carrington event is certainly a benchmark, several space weather events such as those in 1872, 1909, and 1921, have been suggested as comparable in terms of equatorward boundary of auroral visibility (Chapman, 1957; Silverman and Cliver, 2001; Silverman, 2006, 2008). Estimating the equatorward boundary of auroral ovals for these storms supports a feasible comparison of the Carrington event with other extreme space weather events.

Therefore, we first review the evolution of the source active region (AR) on the solar disk at the time. Note that throughout this report we use the terms 'sunspot group' and 'active region' as synonyms. We also recover and examine the contemporary auroral reports in the Russian and Japanese archival material, revise the temporal and spatial evolution of the auroral visibility using known auroral reports (Kimball, 1960; Green and Boardsen, 2006; Humble, 2006; Farrona et al., 2011; Moreno-Cárdenas et al., 2016; Hayakawa et al., 2016, 2018b; González-Esparza and Cuevas-Cardona 2018), and compare them with available magnetograms (Nevanlinna, 2006, 2008; Kumar et al., 2016). With this combined information, we contextualize the results in conjunction with those of the other extreme magnetic storms in observational history (see Chapman, 1957).

## 2. Method:

In this article, we review the contemporary observations of the solar surface and reconstruct the time series of auroral visibility during the stormy interval around the Carrington event. For the observations of the solar surface, we consulted the observational logs by Carrington (1863) and his unpublished manuscripts (RAS MS Carrington 1.3 and 3.2), Schwabe's unpublished observational logs (RAS MS Schwabe 31), and Secchi's reports of his solar observations (OAR MS B13; Secchi, 1859, 1860).

For the auroral visibility, we consulted the observational reports in the yearbook of the Russian Central Observatory (Kupffer, 1860) and Armagh Observatory (see Butler and Hoskin, 1987), newspapers in Portugal, Spain, Australia, New Zealand, and Brazil, and further Japanese diaries and Mexican newspapers (see Supplementary Texts 2.1 – 2.5 in Supporting Information). We then compare them with the known records reviewed in Hayakawa et al. (2018b): reports in contemporary scientific journals (*American Journal of Science* and *Wochenschrift für Astronomie, Meteorologie und Geographie*); ship logs (see Green and Boardsen, 2006; Green et al., 2006); Australian records (see Neumeyer, 1864; Humble, 2006); newspapers in Spain and Mexico (see Farrona et al., 2011; González-Esperza and Cuevas-Cardona, 2018), and East Asian historical documents (see Hayakawa et al., 2016, 2018b). We compute magnetic latitude (MLAT) of the observing sites in the reports, based on the archaeomagnetic field model GUFM1 model covering the position of magnetic dipoles from 1590 to 2000 (Jackson et al., 2000). Note that the canonical archaeomagnetic field model IGRF12 covers the transition of MLATs only after 1900.

We compare recovered records around the Eurasian Continent with the known auroral reports, with magnetic disturbances recorded in the magnetometer in Colaba (Kumar et al., 2016), and those in the Russian Empire at that time (Nevanlinna, 2006, 2008), and also update Figures 3 and 4 of Hayakawa et al. (2018b).

## 3. The Solar Surface:

The storms around the Carrington event occur almost in the maximum of Solar Cycle 10. Figure 1 shows the monthly mean value of the total sunspot number (SSN) (Clette et al., 2014; Clette and Lefèvre, 2016), with two peaks in 1859 October (SSN: 218) and 1860 July (SSN: 222). Likewise, the monthly mean value of the smoothed sunspot area (Carrasco et al., 2016) shows two peaks in 1859 September (2300 msh = millionth of solar hemisphere) and 1860 July (2270 msh). Frequently the sunspot number has two peaks for each cycle, mostly due to the two separated activity maxima of the northern and southern hemispheres (Gnevyshev, 1963; Storini et al., 2002). We contextualize the 1859 storms slightly before the first peak in the SSN and exactly at the first peak in the smoothed sunspot area.

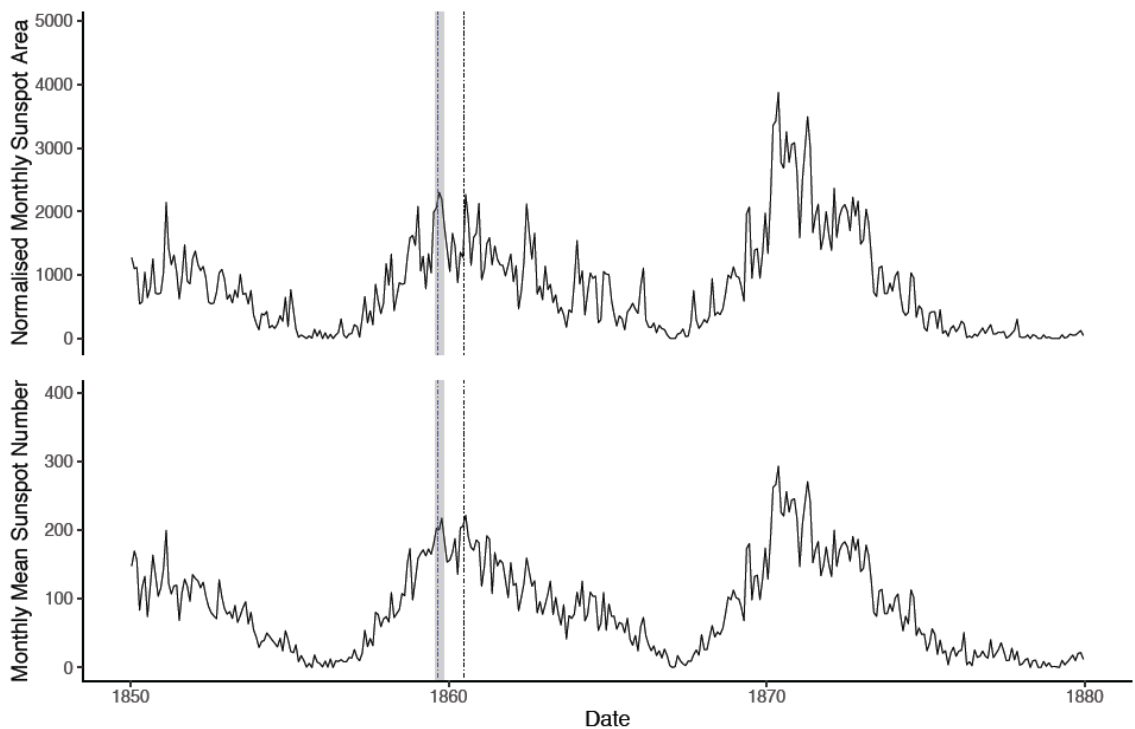

Figure 1: The extreme storms around the Carrington event in 1859 (gray bar) in comparison with the double peak of the monthly mean value of the total sunspot number provided from Sunspot Index and Long-term Solar Observations (SILSO; Clette et al., 2014; Clette and Lefèvre, 2016) in the lower panel and the monthly mean value of the smoothed sunspot area (Carrasco et al., 2016) in the upper panel.

During this enhanced phase near the first peak in Solar Cycle 10, a significantly large and complex sunspot group appeared on the solar disk, which was visible even without a telescope (The Photographic News, 1859, p. 68; *c.f.*, Vaquero and Vázquez, 2009, pp. 57–102; Hayakawa et al., 2017, 2019b). This large sunspot group was monitored by a number of contemporary astronomers such as Secchi, Carrington, and Schwabe. Among them, Father Angelo Secchi was the director of Collegio Romano and a prominent scientist and sunspot observer at that time. He mentioned this group's association with the great auroral display: "It is extremely remarkable that these great perturbations should have coincided with a maximum of solar spots, and should have happened precisely at a moment when an immense spot was visible on the disc of the Sun, even without the aid of the telescope" (The Photographic News, 1859, p. 68).

Within this large sunspot group, Carrington (1859) and Hodgson (1859) witnessed the earliest white-light flare in observational history on September 1. This flare was recorded at 11:18-11:23 UT on September 1, and followed with a synchronized magnetic crochet ≈ 110 nT in the horizontal force at the Earth (Carrington, 1859; Hodgson, 1859; Stewart, 1861; Cliver and Svalgaard, 2004). Note that the source flare for the August storm had not been captured by other contemporary observers (*e.g.*, Neidig and Cliver, 1983; Vaquero et al., 2017). This is not surprising, as the time span of the flare itself in white light is not long and contemporary observers had no concept of flare watch before this discovery.

The exact position and morphology of the sunspot group responsible for these events can be reconstructed from the sunspot drawings and associated observational logs by contemporary observers. In particular, the sunspot group associated with the Carrington flare was recorded not only in Carrington's sunspot drawings but in those by Schwabe and Father Secchi as well. Figure 2 shows Carrington's drawings of a whole solar disk and of the sunspot group that generated an intense flare on September 1. Figure 3 shows Schwabe's sunspot drawings on 1859 August 27 and September 1. Figure 4 displays Father Secchi's solar observations on 1859 August 28 and 31.

Comparison between Figures 2 – 3 shows different viewing aspects in Carrington drawing than in the whole sun drawing by Schwabe. Carrington used a 4.5-inch refractor, which had a focal length of 52 inches, with an equatorial mount and applied a projection method to obtain the solar disk with a diameter of 11 inches for his sunspot observations (Carrington, 1863; Cliver and Keer, 2012). Because of a projected image on a screen, the Sun's north and west are found in the upper and left sides of his original sunspot drawing, respectively. On the other hand, Schwabe used two Keplerian telescopes in which one had a reduced aperture of 1.75 inches and a focal length of 3.5 feet and the other had a reduced aperture of 2.5 inches and a focal length of six feet and dimming glasses to observe sunspots with a direct viewing method (Johnson, 1857; Arlt, 2011). Therefore, his sunspot drawings show a solar image that is reversed in the north-south and east-west directions. The orientation of Schwabe's drawing is in the celestial coordinate system with north pointing down (Arlt et al., 2013).

Secchi's observations were conducted with a Cauchoix achromatic telescope, which had an aperture of 16.9 cm and a focal length of 238 cm, with projection of solar image of diameter equal to 246 mm (Secchi, 1859; OAR MS B13; see also Altamore et al., 2018). In order to make the comparison feasible, we recast the original solar disk drawings in Figures 2 – 4, as seen in the sky on the basis of each observational method.

Comparison between Figures 2 – 4 show that the locations of sunspots in the Schwabe and Secchi's whole disk drawings are consistent with those in the Carrington's drawings.

The sunspot group associated with the Carrington flare is Group 520 in Carrington (1863, p.167), whereas Schwabe (RAS MS Schwabe 31; Figure 3) separated this group to Groups 143 and 142. Note that Schwabe's close-up drawing of Group 143 and Carrington's drawing of Group 520 have been reversed in Figures 2 (Carrington) and 3 (Schwabe). Thus, Schwabe's drawing resembles Carrington's drawing in a general view of the sunspot group. The entire sunspot group may be identified as an Fki-type group in the McIntosh classification (McIntosh, 1990), but the depictions of umbrae are different from each other. This sunspot group is also captured in heliograms at Kew Observatory (RGO 67/266; Figure 5 of Cliver and Keer, 2012) and Secchi's projected sunspot drawings (Group 219 with a smaller group 218 in Figure 4). The sunspot groups recorded in these sources show significantly similar sunspot morphology to those in Carrington's sunspot drawings. Therefore, it is conceivable that the differences in the depiction of umbrae between Carrington and Schwabe stem from difference of their observational methodologies.

These sunspot drawings and heliograms show significantly complex topology of this source AR and indicate strongly mixed magnetic polarity. In theory, the white light (WL) brightenings, which are the footpoints of strong electron beaming from the reconnection site, should be located on both sides of the polarity inversion line. If this is the case, the polarity inversion line crosses the middle of this spot group, indicating a delta-configuration, the most flare-productive category of the sunspots (see Zirin and Liggett, 1987; Toriumi et al., 2017; Toriumi and Wang, 2019).

Schwabe associated the Group 143 with other sunspot groups in early August (127) and early July (112). If this is indeed the case, this group had been extant and recurrent at least for three solar rotations, as is frequently the case with large sunspot groups (Henwood et al., 2010; Namekata et al., 2019). Interestingly, the aurora was reported in China on 1859 August 4 (see Willis et al., 2007) and a negative excursion was recorded in Russia in late September (see Veselovsky et al., 2009), which may support the recurrence of this large sunspot group. Further surveys are required to document the entire lifespan of this AR.

   The original sunspot drawings show that this sunspot group appeared in the eastern limb on 1859 August 25, came across the central meridian around 1859 August 31 and September 1, and went beyond the western limb by 1859 September 7 (Carrington, 1863; Arlt et al., 2013). On September 1, the sunspot was situated at N27.5° – N12.4° in latitude, W28.7° – W6.6° in longitude at ≈ 11.2 h UT (Carrington, 1863, p. 83) and N20.5° – N16.8° in latitude, W24.3° – W8.8° in longitude at ≈ 9.2 h UT (RAS MS Schwabe 31; Arlt et al., 2013), being geo-effectively favourable (*e.g.*, Gopalswamy et al., 2005, 2012; Schrijver et al., 2012).

   The Carrington flare on September 1 was probably preceded by a flare event associated with the August storm. While its onset is not recorded, we may expect it to have occurred somewhere around August 27, assuming a CME transit time of ≥ one day (*e.g.*, Gopalswamy et al., 2005; Lefèvre et al., 2016). The disk center was mostly without sunspots, except for a tiny group (141), situated at N20.9° – N20.7° in latitude, W6.0° – W3.3° in longitude (RAS MS Schwabe 31; Arlt et al., 2013). The only large group (143), which was separated from a small group (142), was situated far eastward then, roughly at E57°, N13° at ≈ 9.2 h UT (RAS MS Schwabe 31; Arlt et al., 2013). It is quite notable that this sunspot group managed to cause a geo-effective ICME even with this unfavourable location (*c.f.*, Gopalswamy et al., 2005; Lefèvre et al., 2016). Cliver (2006) considered the ICME hit the earth "only a glancing below" and was even larger than that of the September storm, assuming the calculated longitude of E55°-E60°.

   Subsequently, another great aurora was reported even down to Athens (Heis, 1861, p. 115; N37°58′, E23°44′, 37.2° MLAT) with a simultaneous extreme magnetic disturbance at Bombay magnetogram ($\Delta H \approx 984$ nT) on 1859 Oct. 12 (see Kumar et al., 2016; Lakhina and Tsurutani, 2017), while the latter seems associated with another sunspot group.

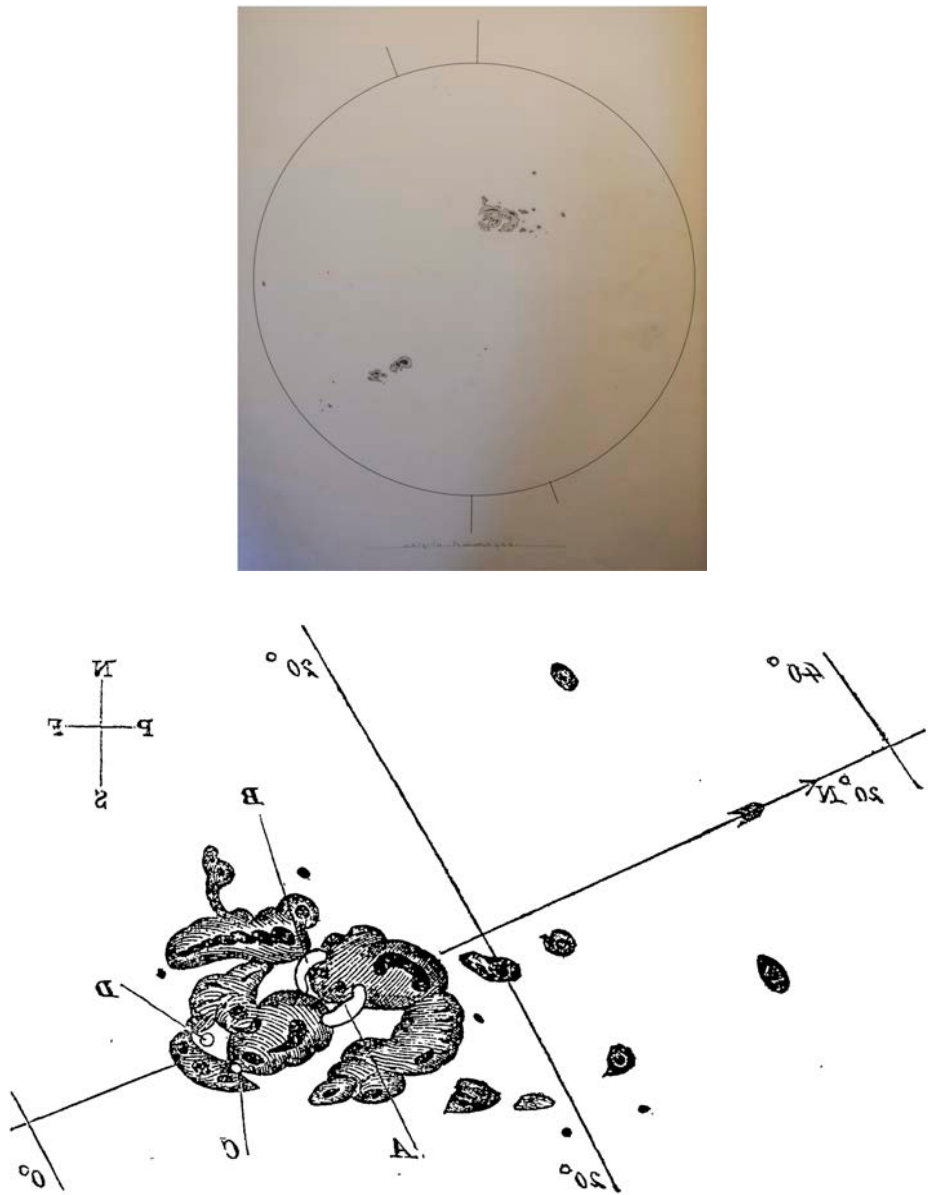

Figure 2: Drawings of a whole solar disk (top) and of the sunspot generated the strongest white-light flares (bottom) made by Richard Carrington on September 1 with its limb enhanced (RAS MS Carrington 3.2, f. 313a; Image courtesy of the Royal Astronomical Society). In both of panels, drawings are reversed from the originals in the horizontal direction as seen in the original solar disk. In the top panel, the Sun's rotational axis is drawn as an oblique line and the sunspot that caused the Carrington flare is in the upper-right quadrant of solar disk.

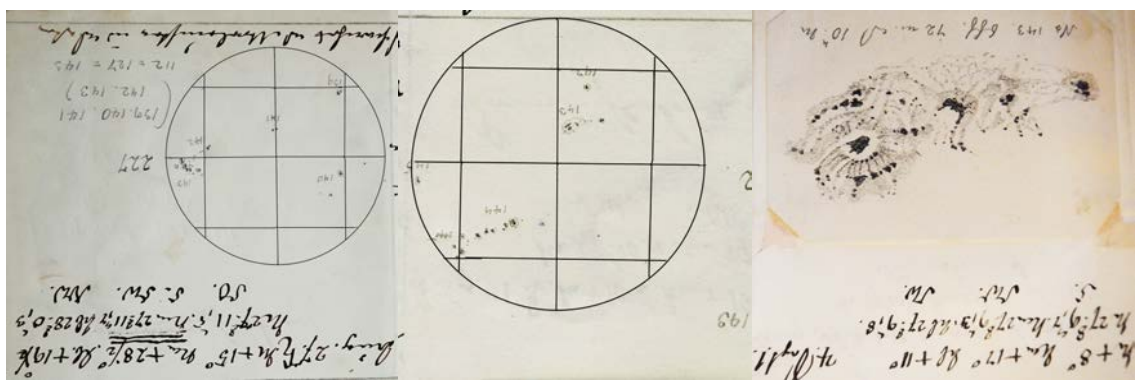

Figure 3: Sunspot drawings by Heinrich Schwabe on August 27 (left), September 1 (centre), and close-up figure of September 1 (right), reproduced from RAS MS Schwabe 31 (p. 131 and p. 136; Image courtesy of the Royal Astronomical Society). Circles in the lower halves correspond to the solar disk, on which the sunspots are drawn with the numbers. The sunspot group that caused the Carrington flare is numbered 143 (on left side of the disk in the left panel, and a little upper right of the disk center in the middle panel). Note that Schwabe separated Carrington's Group 520 to Groups 143 and 142. Close-up drawing in the right panel reveals the details of the Group 143. They are reversed as they were seen on the sky. The solar rotational axis is not shown in these drawings. Their limb and contrast have been enhanced here.

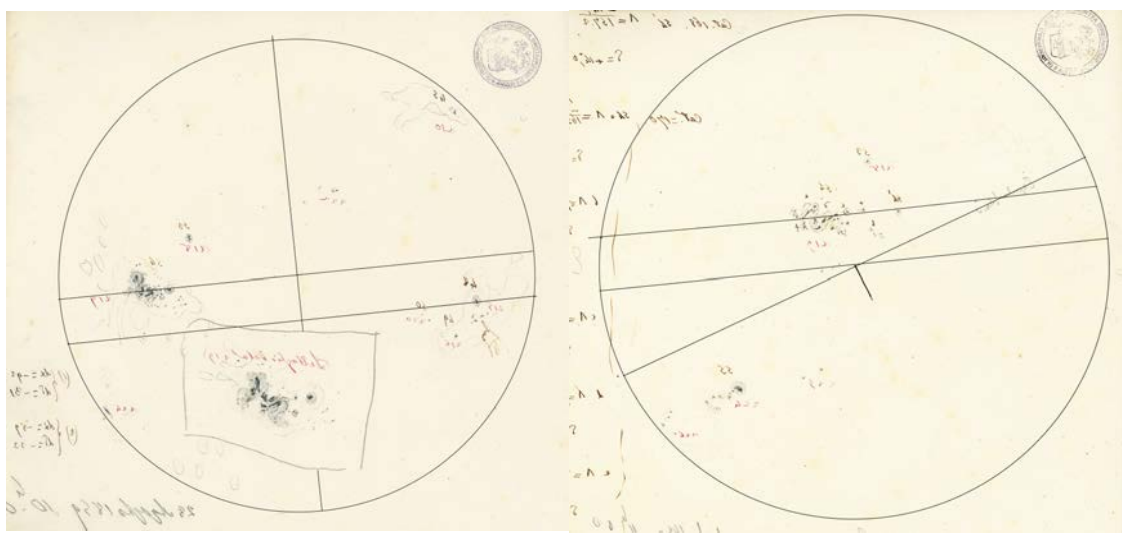

Figure 4: Drawings of the solar disk by Father Angelo Secchi on 1859 Aug. 28 (left) and 31 (right) (OAR MS B13 in Archivio INAF Osservatorio Astronomico di Roma). The left observation includes close-up drawing of the source region of the Carrington event. In both of panels, drawings are reversed from the originals in the horizontal

direction as seen in the original solar disk. The solar rotational axis is shown as a short oblique line in the right drawing. Their limb and contrast have been enhanced here.

**4. Auroral Evolutions and Magnetic Disturbances:**

The large sunspot group (Group 520 in Carrington, Group 143 in Schwabe, and Group 219 in Secchi) caused a series of interplanetary coronal mass ejections (ICMEs) and a subsequent series of magnetic storms and auroral displays between 1859 August 28 and September 4 (Kimball, 1960; Green and Boardsen, 2006; Lakhina et al., 2013; Lakhina and Tsurutani, 2017; Hayakawa et al., 2018b).

Table S1 (see Supporting Information) shows the visual auroral reports around the Eurasian Continent during this time interval, recovered in this article. The Russian yearbook reports auroral displays on August 28/29 at St. Petersburg (56.9° MLAT) and Sitka (59.3° MLAT). The September auroral displays in Russia were seen more widely on September 1/2 – 4/5, throughout the Siberian stations down to Nertschinsk (40.0° MLAT) and Barnaoul (43.2° MLAT). Japanese diaries enable us to add five more auroral reports on September 1/2 that show moderate auroral visibility on the northern coast of Japan. The aurorae were visible down to Hakata (22.6° MLAT), slightly more equatorward than previously known (~23.1° MLAT; Hayakawa et al., 2016).

From Western Europe, auroral records have been recovered in the meteorological records (MS 117) in the Armagh Observatory (see Butler and Hoskin, 1987) and Portuguese and Spanish newspapers. The Armagh records show relatively long auroral visibility during the nights of August 28 and 29 and September 2, 3, and 4. The Portuguese and Spanish newspapers show intensive auroral displays on Aug. 28, which extended even beyond the zenith at Portuguese cities and formed corona aurorae at Lisbon (Pt1: 44.3° MLAT).

From Oceania, we found a series of newspapers in New Zealand and Western Australia. The newspapers in New Zealand reported aurorae mostly on August 29, whereas those in Western Australia reported them on September 2. We also surveyed Mexican and Brazilian newspapers. Consequently, we located two more auroral reports in Mexico. We found no newspapers mentioning auroral observations in Brazil in the database of the National Digital Library of Brazil (http://bndigital.bn.gov.br), while two Brazilian newspapers mentioned auroral visibility at Lisbon (O Cearense, 1859-11-11) and Montreal and New England (Correio da Tarde, 1859-12-10).

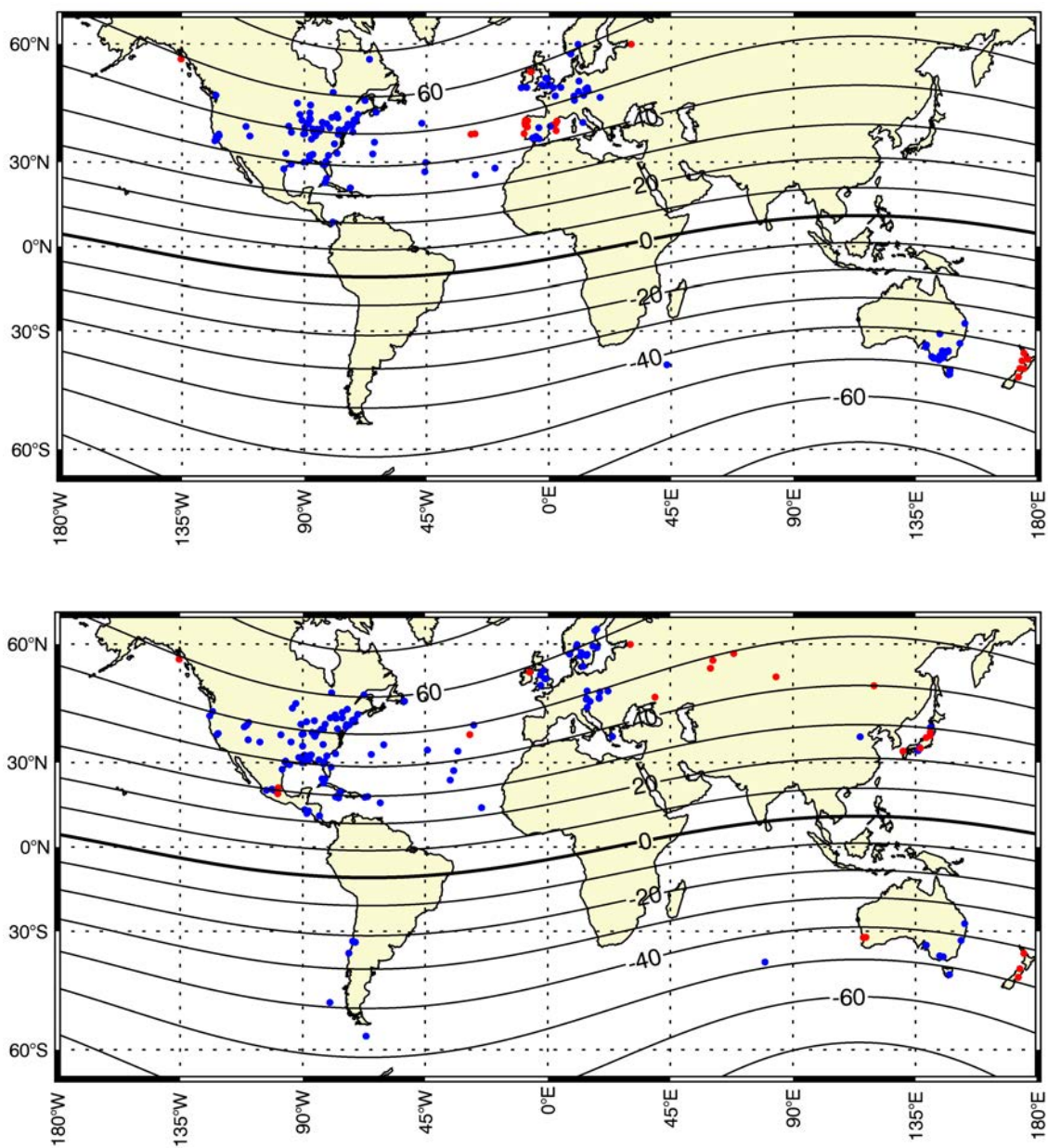

Figure 5: Auroral visibility on 1859 August 28/29 (top panel) and September 1/2 – 2/3 (bottom panel) reconstructed from visual auroral reports. The archival records recovered in this paper are depicted in red colour, whereas the previously known observational sites are depicted in blue colour. The observation at Honolulu is not included here due to its dating uncertainty.

We have integrated these recovered auroral reports from Russian, Irish, Portuguese, Oceanian, Mexican, and Japanese documents with the previously known auroral reports (see Kimball, 1960; Green and Boardsen, 2006; Hayakawa et al., 2018b). Note that the Russian Empire ruled Alaska at that time and had preserved a report at Sitka in Alaska. We have plotted their spatial and temporal extents in Figures 5 and 6. Figure 5 shows the spatial extent of the auroral visibilities on 1859 August 28/29 and September 1/2 – 2/3 on the basis of visual auroral reports from known datasets and new archival records around the Eurasian Continent. This figure shows explicitly that the new data fill the existing gap of observations around the Eurasian Continent, especially in the Eastern Hemisphere (*e.g.*, Kimball, 1960; Green and Boardsen, 2006). These auroral observational sites are partially overlapping with the locations of magnetograms in Russia (Nevanlinna, 2008).

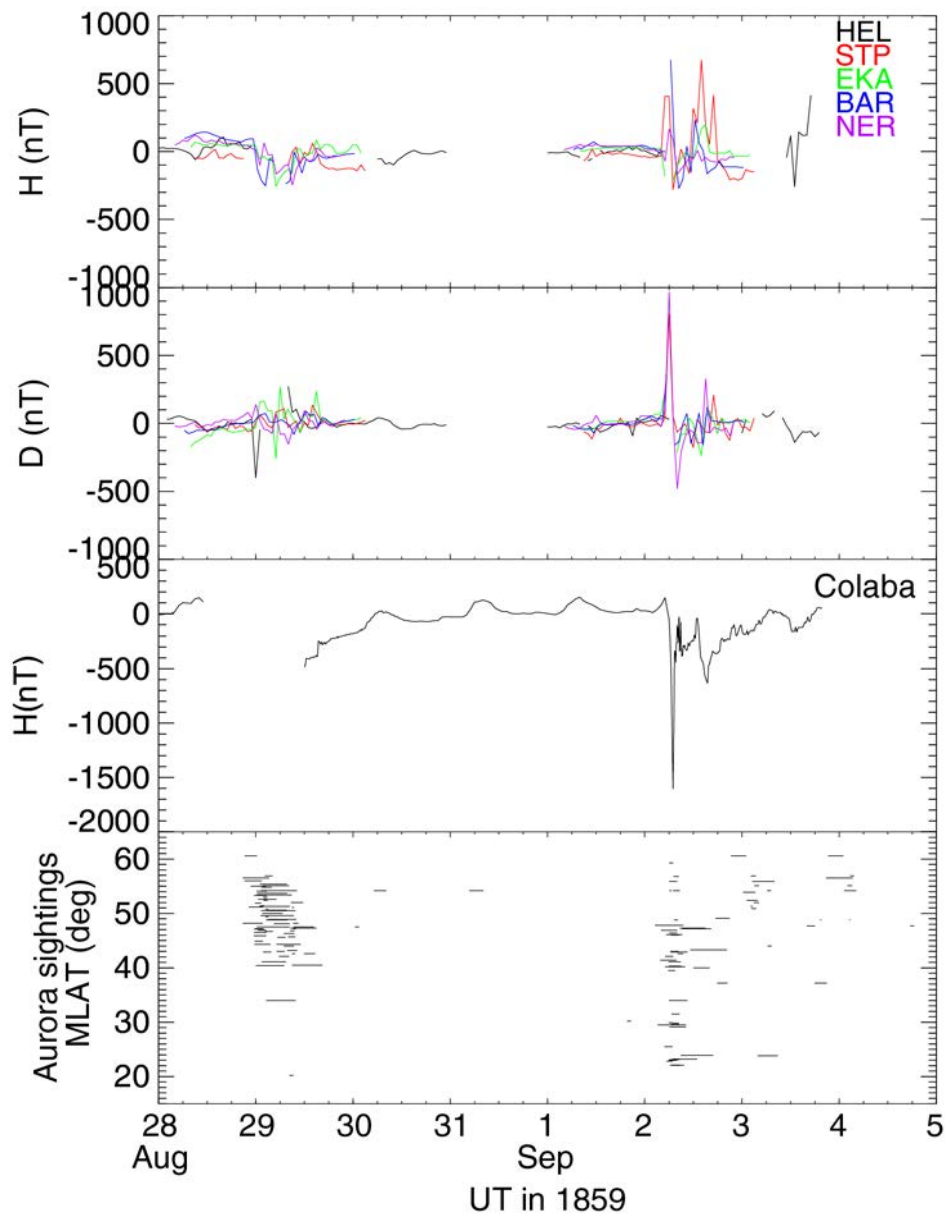

Figure 6: The lower panel is the time series of auroral visibility from August 28 to September 5 with UT on horizontal axis. The corresponding MLAT is on the vertical axis. This time series is compared with the magnetograms in the Russian Empire in the upper panels and that of the Colaba Observatory in the third panel. The abbreviations on the panels for declination (D) and horizontal force (H) in the Russian Empire (the first and second panels) signify observational sites: HEL (Helsinki), STP (St. -Pétersbourg), EKA (Catherinbourg), BAR (Barnaoul) and NER (Nertchinsk).

Figure 6 shows the temporal extent of the auroral visibility during the stormy interval between 1859 August 28/29 and September 4/5. The auroral displays were intermittently visible from August 28/29 to September 4/5, with two remarkable bands on August 28/29 down to ≈ 20.2° MLAT and September 1/2 – 2/3 down to ≈ 20.5°/21.8° MLAT, assuming an auroral elevation up to 400 km (Silverman, 1998; Ebihara et al., 2017). This figure shows the data with a clear start and end of auroral visibility. There are additional reports (not shown) with auroral visibility between these two bands, without a clear description of their start and end. Therefore, some observational records in this interval such as those at Armagh are not plotted in this figure, as they do not have clear description for the start and end of their visibility.

The onset of the first storm is confirmed as a sudden commencement (SC) at 7.5 h UT on August 28 with a relatively large amplitude in the declination of ≈ 30′ with a greater disturbance after 21 h UT (Jones, 1955, p.104). The onset of the auroral visibility is reported roughly after 20.5 h UT on August 28. This is probably because the auroral oval extended more actively in Western Europe and North America and started to be visible after dark there.

The second outburst of auroral displays is almost synchronized with the onset of the sharp negative excursion at Bombay around 4.3 h – 6.7 h UT on September 2. Because this negative excursion falls in the daytime in the Eastern Hemisphere (9.2 h – 11.6 h LMT at Bombay), the auroral displays were mainly reported not in the Eastern Hemisphere but in the Western Hemisphere, such as in the cities along the Caribbean Coast (down to ≈ 22.8° MLAT) and Chile (≈ −21.8° MLAT), with the equatorward boundary of auroral oval around ≈ 30.8° ILAT.

Indeed, aurorae were visible in most equatorward stations immediately after the onset of the magnetic negative excursion (4.3 h – 6.7 h UT). The auroral visibility at Sabine (RG24-2: N11°32′, W083°49′, 23.1° MLAT), St. Mary's (RG24-3: N12°30′, W088°25′, 23.0° MLAT), at a ship in the Atlantic Ocean (WA1: N14°28′, W024°20′, 22.8° MLAT), and Santiago (S33°28′, W070°40′, - 22.1° MLAT) were reported from 6.1 h UT, 5.9 h UT, 6.1 h UT, and 6.2 h UT, respectively (see Table 1 of Hayakawa et al., 2018b). Moreover, if we date the report at Honolulu (20.5° MLAT in visibility and 28.5° ILAT in magnetic footprint) on September 1/2 as in Kimball (1960), the onset of its auroral visibility is 22 h on September 1 in LMT (local mean time) and calculated 8.5 h on September 2 in UT, which is slightly after this negative excursion at Bombay.

This timing may explain why the number of auroral reports from Southern Europe during the August storm is larger than that during the September storm. This negative excursion (4.3 h – 6.7 h UT) falls almost at the end of night in the European sector. Taking Lisbon (N38°43′, W009°08′), one of the westernmost cities in the European sector and hence with one of the sites with the latest sunrise in this sector, as reference, the duration of this negative excursion (4.3 h – 6.7 h UT) was probably affected by twilight and even daylight, as the local sunrise and the local onset of astronomical twilight are calculated as 06:06 UT and 04:33 UT. The Portuguese newspaper at Horta (N38°32′, W028°38′, 47.2° MLAT) confirms this hypothesis describing the auroral visibility from 5.9 UT on Sep. 2 (28 LMT on Sep. 1 in Table S1) "until the dawn light dimmed it" (Pt6).

Nevertheless, the auroral displays remained visible through the recovery phase of the storm and enabled observers in the Russian Empire and East Asia (down to ≈ 22.6° MLAT), and even in Mid Europe, to see these displays into the next night wherever the actual equatorward boundary of auroral oval was at that time.

The magnetogram at Bombay (N18°56′, E072°50′; 10.3° MLAT, E140.5° MLON), whose relative position against the auroral oval was discussed in the context of the possible contribution of ionospheric currents (Green and Boardsen, 2006; Cliver and Dietrich, 2013), is situated in the Eastern Hemisphere and neighbored by the observational sites in the Russian Empire and East Asia (Figure 7). While the visual auroral reports from these sites do not provide records with an elevation angle, we can combine these reports to make conservative estimates for the equatorward extension of the auroral ovals during this stormy interval. The magnetic coordinates of the Siberian station at Nertschinsk (N51°19', E119°36') on September 2 is computed as 40.0° MLAT and W175.5° MLON. This station is situated in a similar magnetic longitude to the auroral observational sites in East Asia such as Inami (HJ2, 23.2° MLAT, W160.4° MLON; N33°49', E135°59'; see Hayakawa et al., 2018b).

In order for aurorae to be visible at Inami up to 10° in elevation angle (see *e.g.*, Shiokawa et al., 1998), the equatorward boundary of the auroral oval needs to be down to, at least, 37.6° ILAT for an auroral height of 400 km. This means the equatorward boundary of the auroral oval was at least extending beyond the zenith of Nertschinsk (40.0° MLAT) during the period of auroral visibility in China and Japan. According to the record at Inami (HJ2), the aurora started to be visible from ≈ 16 LT (≈ 07 UT), which is before sunset. Conservatively, we assume that the aurora actually started to be visible from nautical twilight, that is, ≈ 10.3 UT on September 2.

   As shown in Figures 6 and 7, a positive excursion of the H-component magnetic field at Nertschinsk had ended by the time the aurora started to be visible at Inami. Since Nertschinsk was located in the noon-dusk sector, this positive excursion is probably due to the eastward Hall current flowing in the ionosphere, which is a part of the DP2 current system (Nishida, 1968). It is plausible that the enhancement of the DP2 current system, namely the convection, had just ended by this moment. If so, the enhancement of the convection could have transported cold or warm electrons (Ebihara et al., 2017) deep into the inner magnetosphere to become seed electrons of the aurora. The electrons transported earthward by the convection would remain for a while after the weakening of the convection. The remnant of the electrons is thought to result in the aurora that was visible at HJ2 and HJ8 until ~17 UT on September 2. The bipolar variations of the D-component of the magnetic field at Nertschinsk and STP ($\approx 05 - 10$ UT) are difficult to understand, and will be studied further in the future.

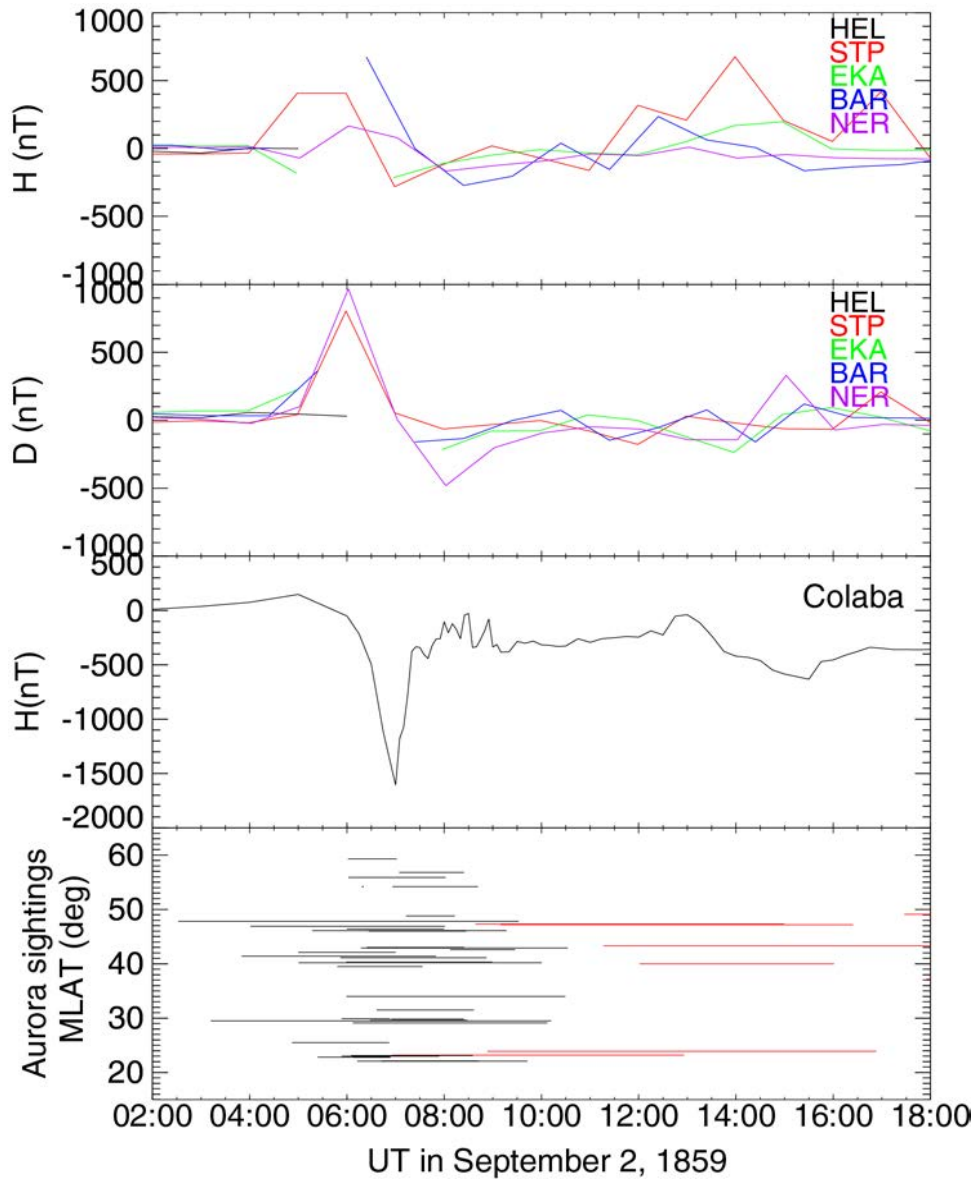

Figure 7: Close-up view of part of Figure 6. The horizontal red lines in the bottom panel indicate the auroral visibility at E072°50′ ± 90° geographic longitude (*i.e.*, in the longitudinal sector centered on Bombay).

**5. Comparison of the Spatial Evolution of the Auroral Ovals for Extreme Events:**
Having presented an updated view of the temporal evolution of the auroral ovals during the stormy interval around the Carrington event, we can categorize the Carrington event not as an exceptionally outstanding event but as one of the most extreme events by comparison with the spatial evolution of the auroral oval for other extreme magnetic storms. Note that the spatial extent of the equatorward boundary of the auroral oval has a good empirical correlation with the storm intensity as indicated by the Dst index (Yokoyama et al., 1998).

During the stormy interval around the Carrington event, the absolute value of the auroral visibility was reported down to ≈ 20.2° MLAT on August 28/29 and ≈ 20.5° MLAT or ≈ 21.8° MLAT on September 1/2. In concert, the equatorward boundary of auroral oval was reconstructed as ≈ 36.5° ILAT on August 28/29 and ≈ 28.5° ILAT or ≈ 30.8° ILAT on September 1/2 (see also Hayakawa et al., 2018b).

The Dst value of the Carrington event is still under discussion (e.g., Tsurutani et al., 2003; Siscoe et al., 2006; Gonzalez et al., 2011; Cliver and Dietrich, 2013). Here, we need to note that, by definition, the Dst value is reconstructed from hourly averages of the horizontal force at four mid-latitude stations. (*e.g.*, Sugiura, 1960; Sugiura and Kamei, 1991). In this sense, the estimates of Dst ≈ −900 (+50, −150) nT on the basis of hourly average of the horizontal force at Bombay better represents the Dst value, although this a single station measurement we still need three more stations for a more standard representation (Siscoe et al., 2006; Gonzalez et al., 2011; Cliver and Dietrich, 2013).

These values are contextualized by comparison with other extreme storms with "outstanding auroras" in 1872 February, 1909 September, and 1921 May (see Chapman, 1957). The equatorward boundary of the auroral oval for the extreme storm on 1872 February 4 is reconstructed as ≈ 24.2° ILAT, based on the reports of overhead aurora up to the zenith at Shanghai (19.9° MLAT) and Jacobabad (19.9° MLAT) (Hayakawa et al., 2018a). The auroral displays themselves are reported down to Shàoxīng (18.7° MLAT; Hayakawa et al., 2018a) and arguably down to Bombay (10.0° MLAT; Chapman, 1957; Silverman, 2008). The magnetogram at Bombay showed the Dst value to be probably < − 830 nT, consistent with a preliminary value from a single station.

Likewise, regarding the extreme magnetic storm on 1909 September 25, the equatorward boundary of the auroral oval is reconstructed as 31.6° ILAT, on the basis of the report from Matsuyama (23.1° MLAT) with an elevation angle of 30° (Hayakawa et al., 2019a). The aurora was also reported from Singapore (−10.0° MLAT), although Silverman (1995) casts doubt on its reliability due to possible contamination from reports of telegraph disturbance. Its Dst value was reconstructed as −595 nT, based on the magnetic observations at Apia, Mauritius, San Fernando, and Vieques (Love et al., 2019a).

Regarding the extreme magnetic storm on 1921 May 14/15 (Silverman and Cliver, 2001; Hapgood, 2019), the aurora was reported down to Apia with a significant magnetic disturbance (Angenheister and Westland, 1921, p.202). The Dst value is computed to be ≈ −907 ± 132 nT, on the basis of magnetograms at Apia, Vassouras, San Fernando, and Watheroo (Love et al., 2019b). The MLAT of Apia is computed as −16.2° MLAT based on the authorized IGRF dipole model (see Thébault et al., 2015). The auroral display was "reaching to an altitude of 22° determined from star positions noted" (Angenheister and Westland, 1921, p.202). Accordingly, we reconstruct the equatorward boundary of the auroral oval as 27.1° ILAT.

These values are comparable to those of the Hydro-Quebec event on 1989 March 13/14, with the most extreme Dst value within the coverage of the official Dst dataset (WDC for Geomagnetism Kyoto, 2015). During this storm, the aurora was visible down to 29° MLAT (Silverman, 2006) and auroral particle precipitation and the auroral electric field were confirmed down to ≈ 40.1° MLAT and ≈ 35° MLAT in the satellite imagery (Rich and Denig, 1992), although that relationship with the visual auroral oval is not completely clear.

Table 1: Comparison of the equatorward boundary (EB) of the auroral oval in absolute value and the Dst values of the outstanding auroras with the Hydro-Quebec Event on 1989 March 13/14, based on RD92 (Rich and Denig, 1992), S+06 (Siscoe et al., 2006), H+18a (Hayakawa et al., 2018a), H+18b (Hayakawa et al., 2018b), H+19a (Hayakawa et al., 2019a), L+19a (Love et al., 2019a), and L+19b (Love et al., 2019b). Note that the Dst value with asterisk (*) indicates a preliminary value using single-station data, due to the availability of complete magnetogram in mid to low latitude (see *e.g.*, Hayakawa et al., 2019a; Love et al., 2019a). The equatorward boundary of auroral oval for the Hydro-Quebec Event is based on auroral particle precipitation and the auroral electric field.

| Event | | | EB of Visibility (MLAT) | EB of Oval (ILAT) | Dst value (nT) | Reference |
| --- | --- | --- | --- | --- | --- | --- |
| Year | Month | Date | | | | |
| 1859 | 8 | 28/29 | 20.2 | 36.5 | ≥ −484* | H+18a |
| 1859 | 9 | 1/2 | 20.5/21.8 | 28.5 / 30.8 | ≈ −850−−1050* | S+06, H+18b |
| 1872 | 2 | 4 | 10.0 / 18.7 | 24.2 | < −830* | H+18a |
| 1909 | 9 | 25 | 10.0 / 23.1 | 31.6 | −595 | H+19a, L+19a |

| 1921 | 5 | 14/15 | 16.2 | 27.1 | −907 ± 132 | This work, L+19b |
| 1989 | 3 | 13/14 | 29 | 35 / 40.1 | −589 | RD92 |

As shown in Table 1, the spatial extent of the Carrington event is comparable to that of other outstanding auroras. As far as currently known, the spatial extent of the equatorward boundary of the auroral oval is most extreme in the 1872 February event (≈ 24.2° ILAT), immediately followed by that of the 1921 May event (≈ 27.1° ILAT) and then the Carrington event (28.5°/30.8° ILAT), while the spatial extent of the Carrington event varies depending on the dating uncertainty in the report from Honolulu (see Hayakawa et al., 2018b).

Given the empirical correlation between the equatorward boundary of the auroral oval and the storm intensity in Dst value (Yokoyama et al., 1998), it seems the Dst value of the Carrington event (September storm) is more likely to be ≈ −900 (+50, −150) nT as an hourly average (Siscoe et al., 2006; Cliver and Dietrich, 2013), comparable to that of the 1921 May storm (see Love et al., 2019b). This comparison tells us that the Carrington event was not the exceptional extreme event, but one of the most extreme events.

Regarding the 1859 August storm, the minimum $\Delta H$ is estimated to be, at least, −484 nT, as the Colaba magnetogram failed to record its main phase on Sunday (Hayakawa et al., 2018b). Colaba was situated on the evening side where the contribution from the ring current is large (Cahill, 1966). Therefore, it is conservatively speculated that minimum Dst was comparable to, or slightly larger than −484 nT.

While the current space weather community expects this kind of event to happen once a century with a potential catastrophe for the modern society (Daglis, 2001; Baker et al., 2008; Hapgood, 2011; Schrijver et al., 2012; Riley, 2012; Riley and Love, 2017; Riley et al., 2018; Dyer et al., 2018), the historical evidence indicates that we need to be slightly more careful about the meaning of 'extreme space weather events'. We were quite fortunate to have the extreme ICME in 2012 July miss the Earth. Some estimates of its potential Dst value appear to be even more extreme than that of the Carrington event (Baker et al., 2013; Ngwira et al., 2013; Liu et al., 2014, 2019). The extremely fast ICME on 1972 August 4 hit the earth with its IMF dominantly northward (Tsurutani et al., 2003; Knipp et al., 2018) causing short-term and local magnetic enhancements that do not appear as part of the Dst record. This storm was very geoeffective even in the absence of a deeply negative Dst value (Knipp et al., 2018). These episodes in the history of space weather indicate that the Carrington event is certainly one of the most extreme events, but is not the single exceptional extreme event.

**6. Conclusion:**

In this article, we have revised the temporal and spatial evolutions of the auroral displays during the stormy interval around the Carrington event. The contemporary sunspot drawings by Richard Carrington, Heinrich Schwabe, and Father Angelo Secchi showed a large and complex source AR between August 25 and September 7. Schwabe and Secchi's sunspot drawings let us confirm the sunspot topology detailed in Carrington's sunspot drawings (Carrington, 1859; Hayakawa et al., 2018b) with their resemblance, even though Schwabe's methodology is different from Carrington's for the solar observation. This resemblance is important because the scientific discussions of the Carrington event have been based on his drawing, and now they are strengthened by these additional contemporary observations. This AR probably caused a significant ICME on August 27 resulting in the first magnetic storm on August 28/29, despite the AR's unfavourable location in the eastern side of the solar disk. The source AR rotated to the disk center on August 31 to September 1 and caused the white-light flare on September 1, associated with the second extreme magnetic storm with low latitude aurorae.

The visual auroral reports from the Russian Empire and Japan enable us to fill the apparent gap of auroral observations in the Eastern Hemisphere, and complimentarily show a long and intermittent auroral visibility through the stormy period from August 28/29 to September 4/5, when compared with the known visual auroral reports. These reports make it possible to estimate that the equatorward boundary of the auroral oval in the Eastern Hemisphere extended at least beyond the overhead positions of the Russian stations ($\approx 37°$ MLAT), even during the recovery phase of the Carrington storms, after the extreme negative excursion recorded at Bombay. The conservative estimate of the equatorward boundary of auroral oval in the Eastern Hemisphere provides further insights on the cause of this magnetic negative excursion in the context of potential auroral contributions.

Revising the spatial evolution of the auroral oval around the Carrington event, we compared the equatorward boundary of auroral oval and Dst value of the Carrington storms with those of the other extreme magnetic storms in 1872 February, 1909 September, 1921 May, and 1989 March. The initial comparison reveals that the Carrington event is probably not the exceptional extreme storm, but one of the most extreme magnetic storms. While this event has been considered to be a once-in-a-century catastrophe, the historical observations warn us that this may be something that occurs more frequently and hence might be a more imminent threat to modern civilization.

**Acknowledgement:**

This research was conducted under the support of the Grant-in-Aid from the Ministry of Education, Culture, Sports, Science and Technology of Japan, Grant Number JP15H05814 (PI: K. Ichimoto), JP18H01254 (PI: H. Isobe), and JP15H05816 (PI: S. Yoden), a Grant-in-Aid for JSPS Research Fellows JP17J06954 (PI: H. Hayakawa), and a mission project of the RISH in Kyoto University. DJK was partially supported by AFOSR grant FA9550-17-1-0258. This project has received funding from the European Union's Horizon 2020 research and innovation program under Grant Agreement No 824135 (SOLARNET). We thank Michael Burton, John Butler, Sian Prosser, Marco Ferrucci and Fabrizio Giorgi, Luís São Bento and Rui Lino, for providing accesses and permissions for researches on historical manuscripts in the Armagh Observatory, the Royal Astronomical Society, INAF Observatorio Astronomico di Roma, Biblioteca Pública e Arquivo Regional João José da Graça at Horta, and Biblioteca Pública e Arquivo Regional Luís da Silva Ribeiro at Angra do Heroísmo. We thank the National Library of Australia, The National Library of New Zealand, Biblioteca Nacional de Portugal, Biblioteca Pública Municipal do Porto, Biblioteca Pública de Braga, Biblioteca Dixital de Galicia, Biblioteca Nacional de España, Arxiu de Revistes Catalanes Antigues, and Hemeroteca Nacional Digital de México for letting us consult newspapers from Australia, New Zealand, Portugal, Spain, and Mexico. We thank Atsushi Soga and Tadanobu Aoyama for their advice on the interpretation of Russian meteorological records, Víctor M. S. Carrasco for providing the background data of Carrasco et al. (2016), Rainer Arlt for his helpful comments on Schwabe's sunspot observations, Christopher J. Scott and Edward W. Cliver for his helpful comments on this article, and SILSO for providing total sunspot number series.